\documentclass{mem}
\usepackage{natbib}\usepackage{txfonts}\usepackage{balance}
\usepackage{graphicx}
\usepackage[a4paper,breaklinks,dvipdfm]{hyperref}
\idline{75}{282}
\begin{document}
\def\teff{$T\rm_{eff }$}
\def\kms{$\mathrm {km s}^{-1}$}

\title{
Radio observations of Supernova Remnants and the surrounding molecular gas
}

   \subtitle{}

\author{G. \,Dubner\inst{1} }

  \offprints{G. Dubner}

\institute{
Instituto de Astronom\'\i a y F\'\i sica del Espacio,
Casilla de Correo 67, Suc. 28,
1428 Buenos Aires, Argentina,
\email{gdubner@iafe.uba.ar}
}

\authorrunning{Dubner}

\titlerunning{SNRs and molecular clouds}

\abstract{
Supernova Remnants (SNRs) are believed to be the main source of Galactic cosmic rays (CR). The strong SNR shocks provide ideal acceleration sites for particles of at least $\sim 10^{14}$ eV/nucleon. Radio continuum studies of SNRs carried out with good sensitivity and high angular resolution convey information about three main aspects of the SNRs: morphology, polarization and spectrum. Based on this information it is possible to localize sites of higher compression and particle acceleration as well as the orientation and degree of order of the magnetic fields, and in some cases even its intensity. All this information, when complemented with the study of the distribution and kinematics of the surrounding interstellar gas, results in a very useful  dataset to investigate the role of SNRs as cosmic ray accelerators. In this presentation, I analyze the radio observations of SNRs and surrounding molecular clouds, showing the contribution of these studies to the understanding of the role of SNRs as factories of CRs.


}
\maketitle{}

\section{Introduction}

A Supernova Remnant (SNR) is the result of the interaction of several solar masses of processed stellar ejecta carrying
  about $10^{51}$ erg of thermal and mechanical energy, with the surrounding circumstellar medium (CSM, likely to be already modified by the progenitor star) or/and the pristine interstellar medium (ISM). Therefore, SNRs consist of the {\it structure} and the {\it products} created during and after the explosion of a star and through the subsequent interaction of the supernova shock front with the surrounding matter. As such they carry unique information about the exploded star, the released nucleosynthesis products, the explosion mechanisms, the circumstellar and interstellar environment where they evolve, and all the physical processes triggered by the sudden injection of a large amount of energy  in the interstellar medium (ISM). 

While the bright Type Ia supernovae (SNe) originate from intermediate to old population low mass C/O stars, more than a half, and probably up to $\sim$ 70\% of the SNRs come from core-collapse SNe of types II and Ib/c \citep[e.g.][]{capellaro99,georgy09},  whose progenitors are young, massive stars. Since massive stars have a short lifetime, many of them explode while they are still inside, at the border, or very close to their parental molecular clouds. Furthermore, massive stars blow strong stellar winds in different evolutionary phases (as OB, LBV or WR stars) creating big bubbles around the star. Therefore, a large fraction of SNe explodes in an environment strongly disturbed by the precursor star. Additionally, the young SNRs have memory of the explosion mechanism and while the remnants of Type Ia explosions tend to be more symmetric, the remnants of type II SNe are distinctly more asymmetric \citep{lopez11}. When the SNRs grow old, it is the surrounding medium which exerts the greatest influence. Therefore,  the shape of intermediate-age and old SNRs usually reflect the impact of the environmental inhomogeneities. Another important difference between Type Ia and Type II-Ib/c SNe, is that the first kind is the result of a thermonuclear collapse that completely destroy the star, while the latter are the product of a gravitational collapse that, at least on theoretical grounds, are expected to leave a compact core (neutron star or black hole) with the subsequent creation of a  pulsar wind nebula (PWN) blown around it. 

After the explosion, blast waves with initial speeds of $\sim$ 5000 to $\sim$ 10000 km s$^{-1}$ sweep up the surrounding material, amplifying irregularities in pressure and density. Strong shocks are driven into the nearby molecular clouds, heating, compressing, accelerating gas, dissociating molecules and creating new ones, and leading to a large variety of physical and chemical processes with different observable effects. 

SNRs have long been proposed to be the natural accelerators of the Galactic cosmic rays, at least up to $\sim 10^{14} - 10^{15}$ eV/nucleon, a fact known as the ``SNRs paradigm''.  SNRs are sources capable to accelerate particles to the right energy through diffusive shock acceleration \citep [for reviews see e.g.,][]{blandford87,jones91,malkov01} and this mechanism indeed predicts acceleration efficiencies in excess of 10\%, as needed. However, all the arguments linking SNRs with cosmic rays are indirect. The best way of proving unambiguously the existence of very-high-energy (VHE) particles, electrons or hadrons, accelerated in SNRs is the detection of VHE (from about 100 GeV up to a few tens of TeV) $\gamma-$rays produced either via Inverse Compton (IC) scattering of VHE electrons off ambient photons or in interactions of nucleonic cosmic rays with ambient matter. 

In the last decade, the emerging field of $\gamma-$ray astronomy has produced numerous discoveries both, in the TeV and GeV energy domains. The vast majority of the newly identified Galactic sources are related to the late stages of stellar evolution, particularly with SNRs and PWNe. However, when confronting observations with theoretical models, there remain a number of important ambiguities and uncertainties from both the observational and theoretical perspectives. In this context, the study of SNRs and their interaction with the molecular clouds is an invaluable tool to advance in the understanding of the many issues involved.

\section{What can be learnt from radio studies of SNRs?}

SNRs are in general good radio emitters of synchrotron non-thermal radiation. Radio imaging at various frequencies and polarimetric studies of SNRs supply information on three important aspects: morphology, magnetic fields and particle acceleration processes. These studies complemented with the knowledge of the distribution, composition and kinematics of the surrounding ISM, can produce a complete picture of the evolution of the SNR and its possible role as cosmic ray factories.

{\bf (a) Morphology:} the shape and brightness distribution is the result of the history of the precursor star, of the explosion mechanism and of the interaction of the expanding shock with the inhomogeneous surroundings. The main drawback of this information is that a three-dimensional expanding body is seen as a bi-dimensional image, and fore- and background features aligned by chance with the SNR can not, in general, be differentiated from proper features of the SNR. One way to overcome this difficulty is by combining radio continuum observations with data from other spectral regimes, like infrared or X-rays, providing complementary information about the emitting plasma. This helps to discriminate whether the observed features are of intrinsic or extrinsic origin. 

The study of the brightness distribution is also a valuable tool to identify PWNe candidates, especially useful for cases where no radio pulsar is detected (for example because it is beaming away). To unambiguously determine if the nebula is the result of the injection of relativistic particles and magnetic fields from a central neutron star, these studies must be complemented with spectral and polarization research of the radio emission confirming  that the spectrum is very flat, with the spectral index $\alpha$ (where S$_\nu \sim \nu^\alpha$)  typically between 0 and $-$0.3, and that the structure is highly polarized.  An example of the use of multifrequency observations as a criterion to decide whether a central nebula is a PWN or not, was recently shown by Giacani et al. (2011)  in the SNR G344.7$-$0.1 located in the vicinity of the unidentified TeV source HESS J1702-420 (Figure 1). In this case the spectral and polarization studies were not conclusive about the nature of the nebula, and from the comparison of the radio emission with the IR emission at 24 $\mu$m and the X-ray emission, the authors concluded that the nebula, though with an appearance suggestive of being a PWN and in spite of the detection of a compact X-ray source nearby, it was in fact the outcome of the SN blast wave encountering dense material along the line of sight.

\begin{figure}[t!] 
\resizebox{\hsize}{!}{\includegraphics[clip=true]{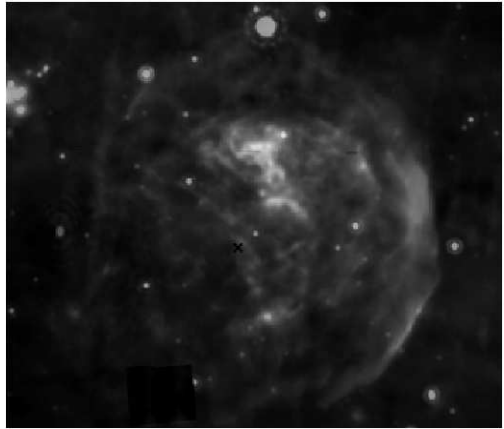}}
\caption{\footnotesize
The true nature of the bright central nebula was established on the basis of the comparison of the emission associated with the SNR G344.7$-$0.1 as observed in radio at 1.4 GHz  and the IR emission at 24 $\mu$m  \citep [from][]{giacani2011}.}
\end{figure}

\begin{figure*}
\resizebox{\hsize}{!}{\includegraphics[clip=true]{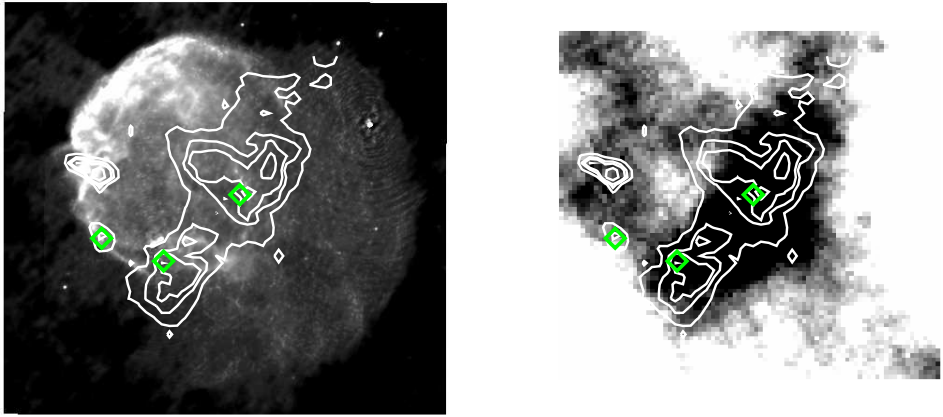}}
\caption{\footnotesize
{\it Left panel:}  VLA image at 330 MHz ($\lambda$90 cm) of the SNR IC~443 as observed by  \citet{castelletti2011}; {\it Right panel:} VERITAS significance map of the TeV $\gamma-$ray source VER J0616.9+2230 in the field of IC~443 
\citep{acciari2009}. In both images, the white contours correspond to the $^{12}$CO:1-0 emission \citep{zhang2010}, while the diamonds mark the location of the OH (1720 MHz) masers \citep{hoffman2003}.
}
\label{eta}
\end{figure*}

{\bf (b) Polarization:} polarimetric studies provide information about how interstellar magnetic fields are amplified and reorganized on relatively small local scales by the expansion of the SN shock. The main disadvantage of this kind of studies is that the polarization is weak (only $\sim$ 10\% of the total intensity or less) and requires long observations to achieve a reasonable signal-to-noise ratio. Also, since linearly polarized radiation passing through  a magneto-ionized medium emerge with its position angle rotated (Faraday rotation), measurements at three or more frequencies are required in order to determine the rotation measure (RM) and obtain the true orientation of the B field. When the radio data are combined with X-ray data, the actual magnetic field strength can eventually be determined. Polarization studies can greatly add to our understanding of SNRs/CR connection, since significant amplification of the magnetic field is expected in SN shocks by CR-induced turbulent processes \citep{bell2001}, allowing for acceleration to the ``knee'' of the CR spectrum, within a SNR lifetime. This effect should be observable through careful polarization studies.

{\bf (c) Radio spectrum:} this is the main instrument to investigate  acceleration processes at the SN shock. The global synchrotron spectrum reflects the average energy distribution among accelerated electrons, while  spatial variations of $\alpha$ across the SNR reveal spatially dependent particle acceleration. In other words, a spectral index map allows localizing the most probable sites of particles acceleration. 

First order Fermi acceleration of relativistic particles at strong adiabatic shocks with a compression ratio of 4 predicts a spectral index $\alpha=-0.5$, though the observed Galactic SNRs show a wider amplitude of spectra, with $\alpha$ varying between $\sim -0.3$ and $\sim -0.8$ \citep{green09}. When synchrotron areas are found to have flat spectral indices, it can be interpreted as a signature of Fermi shock acceleration at sites where stronger post-compression shock densities, accompanied by higher local Mach numbers, and/or higher magnetic field strength result due to impact of the SNR blast wave on a dense cloud \citep [e.g.][]{anderson1993,castelletti2011}, thus confirming that the SNR is accelerating particles at the sites of strong interaction with the surrounding medium. This indication, however, has to be taken with care because flat spectral indices in SNRs can have other origins. In effect,  the presence of ionized gas in the ISM along the line of sight towards the SNR produces the same observable effect on the global spectral index in the low radio frequencies extreme. Interestingly,  as it was first shown by \citet{brogan05}, the presence of localized flat spectrum can also reveal the ionized boundary marking the interface where the SNR is impacting a molecular cloud, turning this into a useful method to recognize elusive SNR/MC interactions. A notable  example of spectral flattening in a SNR originating in at least two different causes, was recently shown by Castelletti et al. (2011)  in the SNR IC~443 (Figure 2). On the basis of the perfect correspondence between a flat spectrum and the infrared emission of multi-ionized species all along the bright eastern border of the SNR, it was concluded that the flat spectrum was produced by the presence of ionized gas. The passage of a J-type dissociative shock impacting a molecular cloud  in the past, dissociated the molecules and ionized the atoms, and these thermal absorbing electrons were now responsible for the observed spectral flattening, while in the interior of the SNR, the flat spectrum features were found to correlate with the location of the densest parts of CO emission (overlapping white contours in Figure 2), revealing the presence of higher shock compression and consequently of particles acceleration.

\section {Supernova remnants and molecular clouds}

Since the first confirmed case of interaction between SNR and dense molecular gas around IC~443 \citep {cornett77, denoyer79}, numerous efforts have been made in observational studies searching for such interactions. Different criteria are applied to demonstrate the existence of physical interaction SNR/MC. Namely, the  search for morphological agreement together with concordance in the distances; molecular line broadening or other type of kinematical perturbation (like the existence of line wings or asymmetries); the presence of OH (1720 MHz) masers,  whose very existence is the strongest evidence of interaction between SNRs and MCs, but whose absence does not rule out it;  high line ratios (for example elevated $^{12}$CO:2-1/$^{12}$CO:1-0 ratio); the detection of NIR $H_2$ lines; etc. However the number of cases of well demonstrated interaction is still small. Jiang et al. (2010) presented a list of SNR/MC interactions compiled in our Galaxy based on the former criteria, finding 34 cases confirmed, 11 probable (based only on morphological agreement) and 19 possible (based on IR colours). Among these 64 cases, 21 SNRs have been reported to have gamma-ray emission. 

As mentioned above, IC~443 is one of the best demonstrated cases of SNR/MC interaction. Figure 2 (right panel) shows  the significance map of the TeV gamma-ray source VER J0616.9+2230, as detected by VERITAS  \citep{acciari2009} compared with  the $^{12}$CO:1-0 emission (white contours) as reported by \citet{zhang2010}, while the diamonds represent the location of the OH (1720 MHz) masers from \citet{hoffman2003}. This SNR has been also detected in the 2-10 GeV range by {\it Fermi}-LAT. The coincidence of the more intense $\gamma-$ray emission with the distribution of the molecular gas, is notable. If the $\gamma-$ray emission originated in CR interactions with the dense external cloud, the dominant emission processes are either bremsstrahlung by relativistic electrons or from neutral pion decay resulting from proton-proton hadronic collisions. Remarkably, most -if not all- the GeV luminous SNRs are known to be interacting with molecular clouds, clearly demonstrating that the interaction with a molecular cloud plays a key role in enhancing the $\gamma-$ray emission  \citep{uchiyama2011}.

\section{Concluding remarks}

One of the essential issues to understand the acceleration processes in SNRs and their role in the production of CRs in the Galaxy seems to be the clear identification of the mechanisms and sites where particle acceleration takes place and the role played by the dense shocked molecular clouds. However, the evidence for interaction SNR/MC is limited and the physical processes involved in the interaction and shock acceleration are far from clear. From the observational point of view, high-quality radio observations of SNRs and of the surrounding interstellar gas, complemented with multispectral information, are the best tool to determine the scenario that can plausibly result in CRs production. 

\begin{acknowledgements}
I am grateful to the Organizers of the conference Cosmic Rays and their Interstellar Medium environment (CRISM 2011) for the kind invitation to participate in the very interesting event. The work benefited from useful discussions with G. Castelletti and E. Giacani. G.D. is a member of CIC-CONICET (Argentina). This work has been partially supported by CONICET, ANPCYT and UBACYT grants from Argentina. 
\end{acknowledgements}

\bibliographystyle{aa}

\end{document}